\documentclass[a4paper]{ESLAB2005}
\usepackage{epsfig}

\begin{document}

\title{The energetic Universe}

\author{X. Barcons\inst{1}}
\institute{Instituto de F\'\i sica de Cantabria (CSIC-UC), E-39005
Santander, Spain}

\maketitle

\begin{abstract}

In this paper I review the main topics on the energetic Universe
that have been put forward as main science goals in the Cosmic
Vision 2015-2025 exercise. I discuss the study of matter under
extreme conditions (both under strong gravity and at ultra-high
densities), the cosmology of baryons (assembly of ordinary matter
in dark-matter dominated structures and the creation of heavy
elements) and the co-eval growth of super-massive black holes and
stars in galaxies along cosmic history. Most of these topics can
be addressed with a large-aperture deep Universe X-ray space
observatory that can be flown soon after 2015, complemented by
gravitational wave observatories (LISA), a focussing gamma-ray
observatory, a far infrared high-sensitivity observatory and an
X-ray survey telescope.

\end{abstract}

\section{Introduction}

Energetic phenomena occur throughout the Universe.  From the
Earth's magnetosphere all the way through the most distant
quasars, not forgetting about the very early Universe, there are
places and times in the Cosmos where particles acquire high
energies. These are often released as high-energy electromagnetic
radiation (X-rays and $\gamma$-rays), which allows us to detect
and study them out to very large distances. Very often, energetic
phenomena are powered by strong gravity fields (around compact
stars and black holes - BH), or very extended potential wells
(groups and clusters of galaxies), strong magnetic fields (neutron
stars) or the combination of various of these phenomena (e.g.,
active coronal stars).

Astronomical observations in the high-energy domain are currently
enjoying a very particular era.  The most powerful X-ray
observatories in orbit launched in 1999, NASA's {\it Chandra}
(\cite{Weisskopf00}) and ESA's {\it XMM-Newton} (\cite{Jansen01}),
have been very recently joined in orbit by JAXA's {\it Suzaku}
(formerly ASTRO-E2). At $\gamma$-ray energies, ESA's {\it
INTEGRAL} is performing smoothly since October 2002
(\cite{Winkler03}). These and other space X-ray observatories are
complemented by a number of ground-based facilities sensitive to
very high $\gamma$-ray energies.

The Cosmic Vision 2015-2025 exercise has provided an opportunity
to revise the most challenging and exciting science that can be
performed in the realm of the energetic Universe in the decade
after next, when the current facilities in orbit have been fully
exploited.  This presentation is an attempt to review some of what
have been considered the most outstanding scientific goals in that
domain for the 2015-2025 timeframe.  These have been grouped under
3 main headings: matter under extreme conditions
(Section~\ref{sec:extreme}, the assembly of baryons in
Cosmological structures (Section~\ref{sec:shape}), and the
evolving violent Universe (Section~\ref{sec:violent}).  Finally I
outline in Section~\ref{sec:future} the main space tools that will
be needed to address these extremely interesting topics in the
2015-2025 timeframe.

\section{Matter under extreme conditions}
\label{sec:extreme}

Testing the intimate nature of matter and of the interactions
among its constituents requires very challenging conditions.
Ground-based particle accelerators, such as LHC at CERN or
Tevatron at Fermilab, are designed and built to probe the
behaviour of matter at very high energies and to probe the
fundamental interactions among particles. The elementary
constituents of matter, now believed to be quarks and leptons,
have been studied to unprecedent detail with such experimental
devices.  So have been the short-ranged weak and strong
fundamental interactions among these particles, as well as the
electro-weak interaction which unifies one of them with the
long-range electromagnetic interaction. In the foreseeable future,
trails of a further unification, that of the electro-weak force
with the strong nuclear one (the one that keeps the quarks bound
inside nucleons), are expected to be seen by colliding matter and
antimatter at the highest energies.

The Universe itself, however, provides us with places where the
environmental conditions can be as extreme, or more, than in any
existing or foreseen laboratory. The fourth of the fundamental
forces of nature, gravity, has a tiny direct effect on the
interactions between elementary particles in a laboratory
experiment, but can be dominant under cosmic conditions. Indeed,
the strongest gravitational fields in nature are those around
neutron stars and black holes (BH).  How does matter behave under
the influence of gravity in the realm of General Relativity, well
beyond the post-Newtonian or any similar perturbative deviation
from Newtonian gravity, can only be studied by observing the
immediate vicinity of black holes and compact stars. Moreover,
nuclear matter inside neutron stars is in such a high density (and
low temperature), thanks again to gravity, that these conditions
cannot be reached in any laboratory.  Although that matter cannot
be directly observed, the physics of strong interactions dictates
its equation of state and therefore the mass and radius of the
neutron star, which are potentially measurable quantities.
Astronomical observations, mostly in the high-energy domain, can
and will provide an important insight into the behaviour of matter
under such extreme conditions.

\subsection{Matter under strong gravity}
\label{subsec:gravity}

\begin{figure}[ht]
  \begin{center}
    \epsfig{file=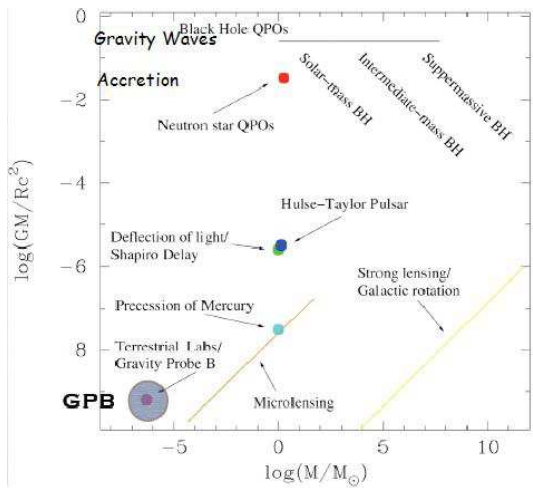, width=7cm}
  \end{center}
\caption{Tests of General Relativity at various field strengths
and mass scales. The strong field limit (top part of the diagram)
can be reached either by observing Gravity Waves or high-energy
radiation from accreting material\label{xbarcons_fig1}}
\end{figure}

General Relativity (GR) is indeed the best ever formulated theory
of gravitation. GR predicts deviations from Newtonian gravity that
have been confirmed and measured very accurately in the weak field
limit. Figure~\ref{xbarcons_fig1} shows the portions of parameter
space spanned by the most representative experiments and tests of
GR. Most of these are concentrated in the weak gravitational
potential limit ($\phi/c^ 2= GM/(Rc^2)<< 1$) and, since most of
them are related to the gravitational field of the Sun or other
stars (the binary pulsar) or the Earth, they are strongly
clustered in the low mass corner.  So far, no deviations from the
predictions of GR have been reported by any experiment, and there
is no reason to believe that GR will not describe gravity
correctly in all the corners of this parameter space.

In the strong field limit, GR predicts a suite of phenomena which
are no longer small perturbations of Newtonian gravity, and which
reflect a strongly curved space-time: strong gravitational
redshift, Lense-Thirring precession, etc. These phenomena need the
gravity of a black hole or a neutron star to be revealed. Strong
variations in these strong gravitational fields will produce
gravity waves that will be the ultimate probes of the behaviour of
space-time closest to the event horizon. Accretion of matter
around a black hole or a neutron star can also probe the curved
space-time within a few Schwarzschild radii. High-energy (X-ray
and $\gamma$-ray) radiation from the innermost regions of the
accreting material can be used to reveal GR effects in the strong
field limit, and to test this theory where its most spectacular
effects are expected. Besides that, black holes span a very wide
range in masses, from a few $M_{\odot}$ in ``stellar'' black holes
to super-massive black holes ($> 10^{5-9} M_{\odot}$) in the
centers of active galaxies (AGN), along with the newly reported
intermediate mass black holes ($10^{2-4}\, M_{\odot}$).

One of the effects of strong gravity on the high-energy emission
from accretion disks is the broadening of X-ray emission lines.
The Fe K$\alpha$ line (at 6.4 keV for neutral Fe), which has been
and will continue to be the best handle for this, likely arises
from fluorescence as the accretion disk is irradiated by the
primary X-ray source. In the innermost parts of the accretion
disk, where most of the power is produced, this line is broadened
by various effects (\cite{Fabian89}), which include the Doppler
effect, relativistic beaming, and gravitational redshift due to
the nearby presence of the black hole. The resulting line profile
is skewed towards softer photon energies due to this last effect.
In AGN the much more distant obscuring ``torus创 can also reflect
primary X-ray radiation, but without any of these broadening
effects.

\begin{figure}[ht]
  \begin{center}
    \epsfig{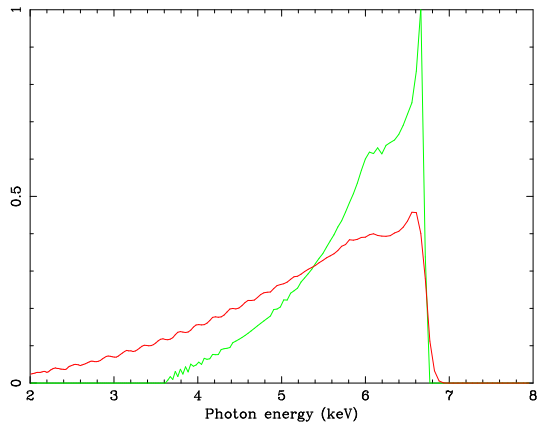}
  \end{center}
\caption{Relativistic Fe line emission profile for non-rotating
(green, more peaked) and maximally rotating black holes (red, less
peaked. \label{xbarcons_fig2}}
\end{figure}

The specific profile of the Fe line depends on many parameters
(disk inclination, among others), but the importance of the low
energy tail is dictated by how close the reflecting material
orbits around the black hole. This, in turn, depends on the spin
of the black hole, because the Innermost Stable Circular Orbit
(ISCO) decreases as the spin parameter of the Black Hole $a=J/Mc$
increases (\cite{Bardeen72}). For a non-rotating (Schwarzschild)
black hole, $R_{ISCO}=3 R_S$, where the Schwarzschild radius is
$R_S=2GM/c^2=3 (M/M_{\odot}) {\rm km}$, but for a maximally
rotating Kerr black hole ($a\sim GM/c^2$) the Fe atoms can orbit
much closer to the black hole $R_{ISCO}\sim 0.5 R_S$. To
illustrate this in simple terms, the closer to the BH that the Fe
line is emitted, the deeper in the gravitational potential well
that X-ray photons will have to escape from and the higher the
fraction of their energy they will have to employ to reach the
observer. Fig.~\ref{xbarcons_fig2} shows the Fe line profile for a
non-rotating Schwarzschild and a maximally rotating Kerr black
hole, assuming that the X-ray emissivity profile is highly
concentrated in the innermost region of the accretion disk
(emissivity $\propto r^{-3}$).

\begin{figure}[ht]
  \begin{center}
    \epsfig{file=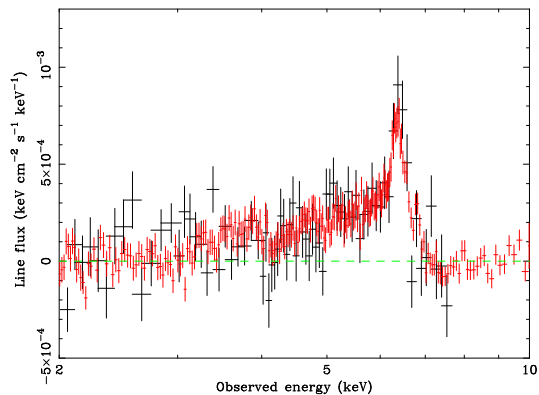, height=7cm, angle=270}
  \end{center}
\caption{Observed Fe line profile for the Seyfert 1 galaxy
MCG-6-30-15 from Young et al (2005). \label{xbarcons_fig3}}
\end{figure}

\cite*{Tanaka95} reported the first clear detection of a
relativistically skewed Fe line profile from the bright Seyfert 1
galaxy MCG-6-30-15 observed with ASCA. Studies of the time
variations of this line (\cite{Iwasawa96}) with the same data,
already indicated that the BH could be rotating.  Higher
sensitivity observations conducted with ESA's {\it XMM-Newton}
(\cite{Wilms01}, \cite{Vaughan04}) confirm the relativistic
profile of the line and clearly call for a rapidly spinning BH
(see \cite{Young05} for a compilation of {\it XMM-Newton} and {\it
Chandra} X-ray spectroscopy on this object).
Fig~\ref{xbarcons_fig3} shows the average X-ray spectral profile
of the Fe K$\alpha$ line for MCG-6-30-15, where it is shown that
the red wing extends down to $\sim 3$ keV, implying $a/M>0.7$.

There are indeed a number of AGN where a relativistically
broadened Fe line has been seen (e.g., NGC 3516 \cite{Turner02}) ,
but in many of the best studied cases the broad red wing is either
absent or difficult to see.  A narrow Fe line, arising from
reflection in the far more distant molecular torus where no
kinematical or relativistic effects are expected, appears to be
ubiquitous (\cite{Yaqoob04}). A combination of both narrow and
broad components is illustrated in the case of Mrk 205
(\cite{Reeves01}). Note also that in the case of MCG-6-30-15
(fig.~\ref{xbarcons_fig3}) the red wing of the Fe line does not
exceed 5-10\% of the underlying continuum in most of the spectrum,
and therefore its detection requires extremely high sensitivity,
something only achievable for the brightest AGN.

The chances of extending the studies of matter under strong
gravity fields to more distant supermassive BH have been revived
by recent work by \cite*{Streblyanska05} and \cite*{Brusa05}.
These studies show that the average X-ray spectrum of distant AGN
and QSOs does show a strong Fe line signal (in fact, implying a
factor 3 overabundance of Fe with respect to solar values), with
broad profile which is reminiscent of a rotating black hole.

Relativistic Fe line profiles have also been seen in Galactic
black hole candidates. Cygnus X-1 (see \cite{Reynolds03} for a
review on the difficulties in detecting this feature) and XTE
J1650-500 (\cite{Miller04}) are among the best studied cases. In
the latter, for instance, there is evidence that reflection occurs
at radii well below the ISCO for a non-rotating BH, implying rapid
spin.

High-energy radiation from accreting material onto BHs is highly
variable. Since the Fe line emission is expected to arise from
reflection, it should follow the variations on the incident
continuum that can also be monitored in line-free regions of the
high-energy spectrum. If X-ray spectral variations could be
tracked down to individual orbit scales, then the whole geometry
of the space-time around the BH could be tested.

Nature is, however, not so simple as revealed in the first attempt
to perform ``reverberation mapping创 on the {\it XMM-Newton} X-ray
data from MCG-6-30-15 (\cite{Vaughan04}). The analysis reveals
that the X-ray spectrum is composed of a roughly constant and
strong reflection component which does not respond to a weaker but
highly variable underlying continuum. \cite*{Miniutti04} propose a
model to explain this that invokes another prediction of GR which
is strong light bending. The fundamentals of that model reside in
the fact that when most of the emission occurs at low latitudes,
the amount of reflection is much larger because the strong BH
gravity bends most of the light from the far side of the BH
towards us.  In such situation we do see a strong reflected
component which should be mostly insensitive to small continuum
variations. When the emission occurs at significantly higher
latitudes, light bending is less important, the reflection
component is weaker, the source is weaker and we should then
expect a nicer correlation between incident (direct) radiation and
the strength of the Fe line. Indeed, specific computations of this
effect should take account of all GR effects at their full
strength.

As it has already been pointed out, the ultimate goal of these
X-ray spectral variability studies should be to go down to
individual orbits.  A remarkable case, where this might have been
already achieved, is that of NGC 3516 (\cite{Iwasawa04}). The time
-averaged X-ray spectrum of this source exhibits a strong Fe line
at 6.4 keV and a weaker redshifted line at 6.1 keV.  A
time-resolved study shows that while the strong blue line is
constant, the red one goes on and off with an apparent periodicity
of 25 ks during a few cycles.  This is what is expected if there
is some feature in the reflecting material rotating very close to
the BH.  Note that this can yield an indication of the mass of the
BH (in this case $\sim 10^7\, M_{\odot}$).  The possibility of
measuring the mass of BHs, by studying spectral variations in
individual orbits is one of the ``holy grails创 in this field.

But orbital motions around BH and NS can also be studied by
accumulating the X-ray light emitted in a broader bandpass,
without going into the details of the X-ray spectral features.
Timing studies of accreting material around BHs and NS reveal the
frequencies of its motion via the power spectrum (Fourier
transform) of the X-ray flux time series. The power spectra
display a superposition of noises (very broad features or
continua) and broad peaks called Quasi-Periodic Oscillations
(QPOs).  The latter are expected to correspond to the proper
frequencies of the motion of the accreting material around the
compact object.  For a stellar BH with mass $10\, M_{\odot}$, and
assuming that emission occurs at the ISCO, the Keplerian frequency
($\nu_K=\sqrt{(GM/r^3)}/(2\pi)$) occurs at $\sim 200\, {\rm Hz}$
and for a NS with mass just above the Chandrasekhar limit ($1.4\,
M_{\odot}$) the orbital frequency exceeds 1 kHz. Therefore, msec
timing is the key ingredient to characterize motions around $<
10\, M_{\odot}$ compact objects.

The solution of the equations of motion in a Kerr spacetime
predicts 3 fundamental frequencies, as the GR motion does not
occur in a plane, nor does it describe closed orbits. These
frequencies are called {\it orbital} $\nu_{\phi}$ (which in the
absence of rotation of the central object reduces to the Keplerian
frequency), {\it radial} $\nu_r$ and vertical $\nu_{\theta}$
epicyclic.  All 3 frequencies are equal to the Keplerian orbital
frequency in Newtonian mechanics, but in GR $\nu_r$ is different
-independent on the spin parameter of the Kerr metric-, and
$\nu_{\theta}$ also differs from the orbital frequency if the
compact object rotates. This leads to several predictions. First,
regardless on whether the compact object rotates or not, there is
a periastron precession with frequency
$\nu_{peri}=\nu_{\phi}-\nu_r$, as in the well studied classical GR
test of Mercury.  If the compact object rotates, then
$\nu_{\phi}\neq\nu_{\theta}$ and there is Lense-Thirring
precession of the orbital plane (i.e., the plane of the orbit
precesses around the spin of the compact object) with nodal
frequency $\nu_{nodal}=\nu_{\phi}-\nu_{\theta}$.

The very detection of the ISCO around a BH or a NS is in itself a
confirmation of General Relativity.  If the mass of the compact
object were known, its spin could then be deduced.  kHz QPOs have
been indeed detected in a number of Galactic Black Hole candidates
and Neutron Stars (particularly with NASA's {\it RXTE}
observatory, \cite{Bradt93}), but the interpretation of the
various peaks in the Fourier spectrum is not straightforward and
appears complicated. Relations involving the width of the various
frequency peaks ($\Delta\nu$), their coherence
($\nu_0/\Delta\nu$), etc. need to be used in the framework of
different models to understand the full QPO spectrum
(\cite{vdKlis04}). Every model (relativistic precession,
relativistic resonance, frequency beating etc.) predicts a number
of distinctive imprints in the power spectrum, particularly at low
frequencies. There is general belief that this technique can
provide a detailed insight on the motions of the accreting
material in the strongly curved space-time around a BH or a NS.
The best approach would be, however, to have the possibility to
study orbits individually, i.e., to have a very high-throughput
X-ray detector capable of reliably measuring high-accuracy fluxes
every fraction of a msec orbit.

\subsection{Matter at supra-nuclear densities}
\label{subsec:density}

Neutron stars are amongst the densest objects in the Universe,
with their core density being 5 to 10 times larger than an atomic
nucleus.  The physics of matter at these densities is largely
unknown. From the phenomenological point of view, laboratory
ion-collision experiments can only partially approach the
environmental conditions in the core of a NS, because the
different ``temperature创 and also the different proton fraction
(typically very small in NS). Uncertainties in the symmetry energy
function in the energy of a nucleonic system are already large at
nuclear densities (it combines a term scaling with nuclear mass
and another one with the surface which are difficult to
disentangle), and extrapolations at significantly higher densities
make the situation even worst. For a review see
\cite*{Lattimer04}.

\begin{figure}[ht]
  \begin{center}
    \epsfig{file=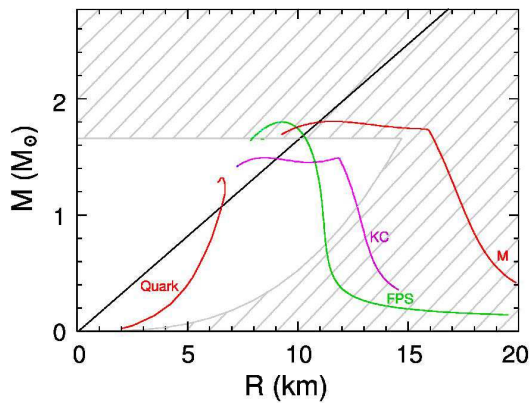, width=7cm, angle=0}
  \end{center}
\caption{Mass-Radius parameter space for Neutron Stars, along with
predictions from various EoS. The dashed region is excluded by
assuming the detection of orbital motion at $\nu_{orb}= 1 \, {\rm
kHz}$. The diagonal line corresponds to a surface redshift of
 $z_{grav}=0.35$. \label{xbarcons_fig4}}
\end{figure}

From the theoretical point of view, the situation is not any
better.  The ``classical创 composition of a NS with a majority of
neutrons and a tiny fraction of protons (with the corresponding
neutralizing electrons and muons) has many chances of being too
simplistic at supranuclear densities. The core of a NS could be
well rich in pion or kaon condensates or it could consist of a
soup of unconfined quarks (see \cite{Lattimer01} for a collection
of models). Even more, it could consist of "Strange" quark matter,
i.e., a combination of the quarks $u$ (up), $d$ (down) and $s$
(strange) with a considerably different set of properties.  Each
one of these possible compositions leads to a different equation
of state (EoS). While it is not possible to probe directly the EoS
at these densities, the mass and radius of the resulting NS is
strongly affected by it. Figure~\ref{xbarcons_fig4} shows a
selection of Mass-Radius relations for a number of equations of
state, where it is seen that "Strange" stars predict significantly
smaller masses than "normal" (with $u$ and $d$ quarks only) stars.

NS masses can be best measured in close compact binaries, such as
the binary pulsar PSR 1913+16, where Shapiro delay or orbital
period variations due to the loss of energy by Gravitational
radiation, can disentangle individual masses. In this case, the
mass is $1.44 \, M_{\odot}$. Causality (the speed of sound has to
be less than the speed of light) sets up an upper limit to the NS
mass at around $3\, M_{\odot}$.

As far as measuring NS radii are concerned, the situation is more
difficult, with the possible exception of NS of known distance and
under the assumption of a Stefan-Boltzmann (black body) emission
law. There are, however, X-ray observations that can help
measuring $M$ and $R$ for neutron stars. Following with the timing
analysis discussed under \ref{subsec:gravity}, the detection of
the frequency of orbital motion $\nu_{orb}$ from kHz QPOs places
useful constraints on the $M$-$R$ parameter space. This comes
simply from the fact that $R_{orb}>R_{ISCO}$ and that
$R<R_{ISCO}$. In fig.~\ref{xbarcons_fig4} we display the region
excluded, assuming that orbital motion has been detected at 1 kHz.

An alternative method to constrain this diagram, even for isolated
neutron stars, has been put forward by \cite*{Cottam02}.  The NS
EXO 0748-676 occasionally bursts material, and the photospheric
absorption lines in its X-ray spectrum are affected by the strong
gravity at the NS surface.  By adding the X-ray spectra of 26
bursts, \cite*{Cottam02} were able to measure a gravitational
redshift of $z=0.35$ at the surface of this NS.  This implies a
roughly constant value for $M/R$ which is also displayed in
fig.~\ref{xbarcons_fig4}.

More detailed information on the EoS of matter at supranuclear
density could be gained if X-ray spectra could be obtained with
very high signal to noise and at high spectral resolution, since
the Stark effect due to the electric field at the NS surface would
broaden the photospheric absorption lines with FWHM $\propto
M/R^2$. Detection of this pressure broadening would break the
degenracy between $M$ and $R$ set by the gravitational redshift
and measure $M$ and $R$ to sufficient accuracy to single out an
EoS.

\section{The Cosmology of baryons}
\label{sec:shape}

After years of experiments and observations, a consensus has been
reached among cosmologists about the basic ingredients of the
Universe. The Universe began some 14 billion years ago in a big
event (the Big Bang) where the ordinary laws of physics do not
apply.  The first direct electromagnetic radiation that we receive
from the past of the Universe is the Cosmic Microwave Background
(CMB), which results from the recombination of electrons and
protons to form atoms, when the universe was less than half a
million years old.  CMB maps reveal an extremely uniform Universe
at that epoch, but with small wiggles which were the seeds of the
large-scale structures that we see today. How the Universe went
from an extremely smooth phase to the highly structured situation
we see today, with clusters, superclusters, voids, filaments and
all sorts of structures, is the main goal of Astrophysical
Cosmology.

Today, ordinary, baryonic matter represents 4-5\% of the total
content of the Universe, and almost half of it is in an unknown
location.  About 23\% is made of Dark Matter (DM), which binds
galaxies and clusters via gravitational attraction (the only
manifestation of DM so far).  The bulk of the Universe is
contributed by an even more exotic, extremely uniform component,
Dark Energy (DE), which in fact shapes the geometry of the
Universe and acts as a peculiar ``repulsive创 force. The relative
amount of DM and DE changes with cosmic time, and it is expected
that in the past the Universe was DM-dominated rather than
DE-dominated as it is today. DM is approximately pressureless, but
DE has, to first order, the pressure of an unstable vacuum
$p_{DE}=-\rho_{DE}c^2$, which gives it this particular
``accelerating创 character in the history of the expansion of the
Universe.

In principle, only the 4-5\% of the baryons is all we can observe
via electromagnetic radiation. The DM potential wells of groups
and clusters of galaxies, the largest gravitationally bound
structures in the Universe, have virial temperatures which imply
X-ray temperatures for the baryons trapped in them.  Indeed,
groups and clusters are X-ray emitters.  Tracing the history of
the assembly of baryons into these large-scale structures, finding
the almost 50\% of them which are missing (probably in a warm/hot
intergalactic medium), and studying when and where the heavy
elements of which the current Universe is made of were produced
along cosmic history are amongst the most important goals in what
can be called the "Cosmology of ordinary matter".  X-ray and
$\gamma$-ray radiation are the best handles towards that goal.

\subsection{Birth \& growth of galaxy clusters} \label{subsec:clusters}

The gravitational attraction produced by DM is the prime actor in
the assembly of groups and clusters. Gravity binds the baryons
together and heats the intra-group or intra-cluster gas to high
temperatures.  However, this is not the end of the story, as
gravitational heating would predict, for example, a relation
between the X-ray gas luminosity and temperature $L_X\propto T^2$,
while a significantly steeper scaling law ($L_X\propto T^3$) is
observed (\cite{Arnaud99}). The ``entropy floor创 discovered in
groups of galaxies (e.g., \cite{Ponman03}), and unlikely to be
produced by an overall pre-heating process, is another
manifestation of the complexity of the problem.

Gas cooling at the cluster centers, heating by Supernovae or AGN
are ingredients that surely come into play in determining the
structure of groups and clusters. Thanks to {\it XMM-Newton}, it
is now known that cooling of the baryons in cluster cores is not
so dramatic as predicted by the older cooling flow models (e.g.,
\cite{Peterson03}), with little gas cooling below 2 keV.
\cite*{Fabian03} argue in the case of the Perseus cluster that the
continuous blowing bubbles of electron gas by the central radio
AGN, can balance the cooling within the cluster core.

To understand how groups and clusters form, it is first necessary
to disentangle the role of the various processes that affect the
cluster entropy (cooling, Supernova and/or AGN heating) for a
range of cluster masses and at various epochs of cosmic history.
Obtaining temperature and gas density cluster profiles spanning a
wide range of cluster mass and redshift is the way to go. The
detailed physics of baryons in cluster cores will in addition
require high-spectral resolution, spatially resolved X-ray
spectroscopy of cluster cores.  Last, but not least, the role of
turbulence, high energy tails (discovered in several clusters of
galaxies), cosmic rays (of which clusters are full), magnetic
fields and other energetic phenomena will also require
$\gamma$-ray observations of clusters. Most of these are beyond
the capabilities of present X-ray and $\gamma$-ray
instrumentation.

Once it is known how clusters work, there is the possibility of
using them as cosmological tools, being the most massive
gravitationally bound structures.  The population of clusters is
indeed very sensitive to the values of cosmological parameters
(\cite{Griffiths04}). The DUO (Dark Universe Observatory) mission
proposal, showed that by studying the number of clusters as a
function of redshift, one can derive not only the amount of DM and
DE, but also the equation of state of DE.  A further crucial piece
of information can be obtained from the spatial distribution of
clusters, by sampling sufficiently large contiguous areas of the
sky.

\begin{figure}[ht]
  \begin{center}
    \epsfig{file=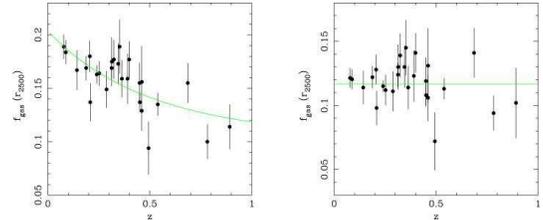, width=7cm, angle=0}
  \end{center}
\caption{Gas fraction of clusters as a function of redshift for a
CDM cosmology (left) and a concordance cosmology (right), adapted
from Allen et al (2004). \label{xbarcons_fig5}}
\end{figure}

Further information on the Cosmological parameters can be gained
by studying the cluster gas fraction for large and relaxed
clusters as a function of cosmic history, as recently discussed by
\cite*{Allen04}. The gas fraction (baryon to DM mass) in these
objects is supposed to be representative of the full Universe on
average and therefore should be constant along cosmic history.
Estimating both the gas mass and the DM mass from observable
quantities involves the use of the luminosity distance which in
turn depends on the Cosmological parameters.  Current studies are
very short on the amount of high-redshift clusters, but already
indicate that the concordance cosmological parameters discussed
above give a much better fit than other cosmological parameters,
for example a standard Cold Dark Matter one (see
fig.~\ref{xbarcons_fig5}, from \cite{Allen04}).

\subsection{The missing baryons}
\label{subsec:WHIM}

The baryonic component of the Universe is well restricted by
several Cosmological tests (including primordial nucleosynthesis,
\cite{Kirkman03}  and CMB anisotropies, \cite{Bennet03}) to be
around 4.5\%. Lyman-$\alpha$  clouds (including damped
Lyman-$\alpha$ absorption systems) detected as HI Lyman-$\alpha$
absorption lines towards distant QSOs are seen to dominate the
baryon content of the Universe at high redshift ($z>2$,
\cite{Storrie00}). At lower redshifts the number density of
Ly$\alpha$ absorbers and the subsequent contribution to the
ordinary matter content of the Universe decrease.
\cite*{Nicastro05} estimate that the total budget of baryons is
locally 2.5\% of the total content of the Universe, with a further
2.1\% (i.e., about half of the total amount of baryons) missing.

Detailed simulations of the cosmological evolution of baryons
invariably show that the Lyman-$\alpha$  absorbing gas at
temperatures $\sim 10^4$ K undergoes shock heating at lower
redshifts and its temperature rises to $10^{5-7}\, {\rm K}$.
According to these simulations, baryons in this warm and hot
intergalactic medium (WHIM) could probably account for the missing
fraction of the baryon budget in the local Universe. These baryons
are expected to be distributed following filamentary structures
dictated by the underlying DM distribution.

Given the sparsity of these baryons in the WHIM, they would be
best seen via resonance absorption lines of highly ionised species
(OVI, OVII, OVIII, NeIX, etc.) towards bright background sources
(typically AGN). Most of these lines occur in the soft X-ray
regime, and therefore sensitive high-resolution X-ray spectroscopy
is the best tool for this purpose. {\it Chandra} and {\it
XMM-Newton} have already started the run to detect absorption
lines from the WHIM, with a handful of positive hits - many of
them arising in local gas (\cite{Nicastro02}, \cite{Rasmussen03}).
\cite*{Nicastro05} compute that, within (large) errors, the mass
contained in the WHIM is consistent with the missing fraction of
baryons.

The observations needed to detect these absorption lines are at
the very limit of contemporary X-ray instrumentation.  Both larger
effective area and better spectral resolution are needed to sample
a large number of lines of sight to improve the statistics and to
trace the filamentary structures predicted by the simulations, as
well as to reach significant redshifts to test how baryons in the
intergalactic medium are heated towards the current epoch.

\subsection{The creation of heavy elements}
\label{subsec:elements}

The heavy elements that constitute the Universe today, were
produced in stellar cores and dispersed in Supernova explosions.
Locally, the detection of elemental abundances in Supernova
Remnants (SNRs) is the most direct way to study this process.
Spatially resolved X-ray spectra of SNRs show in detail how the
various elements are propagated into the interstellar medium.
Nuclear lines observed in  $\gamma$-rays can also determine
elemental abundances, in particular of heavy rare elements.

Beyond our Galaxy and perhaps a few more in the Local Group,
intracluster gas is probably the best tracer of heavy element
abundances as a function of redshift. X-ray spectra of clusters
are rich in emission lines superimposed to thermal bremsstrahlung.
From the line emission, elemental abundances of a variety of
elements can be derived, and related to the history of star
formation. The Fe abundance is easier to obtain, as the K$\alpha$
complex at 6-7 keV is usually strong and isolated from other
spectral features.  Obtaining abundances of other elements (Mg, O,
Si, etc.) needs observing at softer X-ray energies, which would
usually require higher resolution spectroscopy.

One of the most intriguing results found in this area is that the
Fe abundance in clusters stays approximately constant at 0.3 of
the solar value, out to the highest redshifts ($z\sim 1.1$,
\cite{Hashimoto04}).  This means that heavy elements are already
in place at these early epochs, and therefore most of the
enrichment in heavy elements has happened before. However, the
number of clusters at $z>1$ is very small, and the detection of
line emission from them is at the limit of the capabilities of
current instrumentation. To trace the history of heavy element
enrichment more sensitive X-ray observatories equipped with
high-resolution spectrometers are needed.

\section{The evolving violent Universe}
\label{sec:violent}

One of the very first discoveries of X-ray Astronomy was the
Cosmic X-ray Background (XRB, \cite{Giacconi62}). This energetic
radiation that fills the Universe is known today to be the
integrated radiation produced by accretion onto supermassive black
holes (SMBH) along cosmic history. These grown SMBHs are those
that we see today in the centers of virtually all galaxies and
that comprise $\sim 0.4\%$ of their bulge mass
(\cite{Ferrarese05}). According to the AGN unified models for the
XRB (\cite{Comastri95}), most of this accretion occurs in obscured
mode (more than 50\% of it, \cite{Fabian99}), and therefore its
direct detection can only be achieved in hard X-rays.

This general qualitative picture of the XRB being the echo of the
growth of the supermassive black holes opens a number of
questions. The first one is how SMBHs (or their seeds) were born
and whether they can be detected or not at the time of birth. The
second one is how they grow from their probably small initial mass
to their very large masses that we see today ($>10^9\,
M_{\odot}$).  Finally there is the question on how the birth and
growth of SMBHs relates to the formation of galaxies and their
stars.  We have now clear clues that there is a link between both
processes, but how exactly they work is not yet understood.

There are a number of additional questions regarding the evolving
violent Universe, some of which might certainly be related to the
topics just discussed.  One of the outmost importance is the
nature of Gamma Ray Bursts, and whether any of them (perhaps yet
to be discovered) are related to the birth of SMBHs. This and
other topics will certainly meet progress within the Cosmic Vision
2015-2025 timeframe.

\subsection{Birth \& growth of supermassive black holes}
\label{subsec:SMBH}

Numerical simulations show that the very first stars that formed
in the Universe grew up from a seed of about $\sim 1 M_{\odot}$ to
a few hundred solar masses within a few million years
(\cite{Abel02}). These stars exploded leaving a BH of mass of a
few $\sim 10\, M_{\odot}$, and sterilizing a large region ($\sim
10^6\, M_{\odot}$) for further star formation around them.
According to the numerical simulations, these seed BHs left the
scene at a relatively large velocity ($\sim 10\, {\rm km}\, {\rm
s}^{-1}$) and it is therefore unclear whether all of them were
able to start any efficient accretion process.

These first small BHs that formed early on in the history of star
formation, could well be the seeds of their grown-up version that
we see in the centers of galaxies today. For this to happen, they
need to accrete matter rapidly, in an almost exponential fashion
(\cite{Archibald02}).  Massive BHs ($\sim 10^8\, M_{\odot}$) are
already in place in the most luminous QSOs found at early epochs
($z > 4$). This means that by $z\sim 10$ the first mini-QSOs,
hosting a BH of mass $\sim 10^4\, M_{\odot}$, should be there
accreting close to the Eddington limit.

\begin{figure}[ht]
  \begin{center}
    \epsfig{file=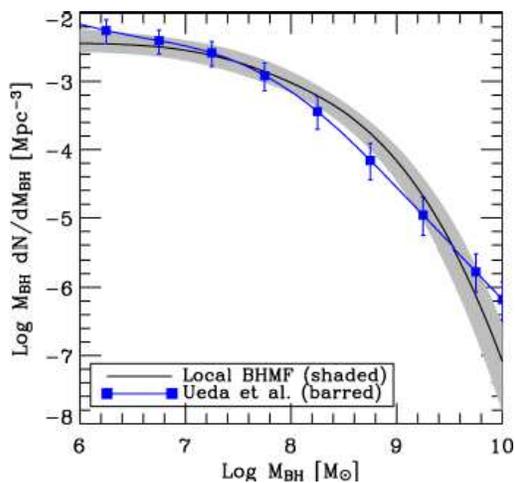, height=7cm, angle=270}
  \end{center}
\caption{Local SMBH function as observed from galaxy kinematics
(dots) and as computed by assuming efficient accretion growth in
AGN (adapted from Marconi et al 2004). \label{xbarcons_fig6}}
\end{figure}

Copious X-ray radiation, and in particular in hard X-rays for
obscured objects, is emitted during the growth phase by accretion
of the SMBHs. X-ray deep surveys (see review by \cite{Brandt04})
are then the best tool to characterize the first stages of the
growth of black holes by accretion.  Indeed, other processes might
also be important in the growth of black holes, namely merging or
tidal capture.  To first approximation and to the best of today's
knowledge, highly efficient (probably requiring rotating Kerr BH)
accretion is probably the dominant process (\cite{Marconi04}) in
the ``growth创 of SMBH, as it can match the AGN X-ray luminosity
functions (the output from accretion) to the local SMBH density
(see fig.~\ref{xbarcons_fig6}). The birth of SMBHs is not well
understood. The galaxy merger rate peaks at $z\sim 2$, which is
where the unabsorbed type 1 AGN population also peaks, implying
that mergers might be important. However, the role of type 2 AGN,
whose population peaks at significantly later epochs, is not
known. Whether these are two distinct unrelated populations, or
follow some sort of evolutionary sequence is an open and debated
question.

The XRB remains a key handle to quantify the amount and ``mode创
of accretion.  Most of its energy density resides at $\sim 30\,
{\rm keV}$ (see, e.g., \cite{Fabian92} for a review), implying
that an important fraction of the energy generated by accretion
onto SMBHs is heavily obscured.  Current instruments ({\it
XMM-Newton}, {\it Chandra} and {\it Integral}) are not sensitive
enough at these energies, and therefore not much can be said
beyond extrapolations. Sensitive instruments at these hard
X-ray/soft $\gamma$-ray energies are needed.

\subsection{Supermassive black holes and star formation} \label{subsec:coeval}

The birth and growth of SMBHs in the centers of galaxies cannot be
independent of the birth and growth of the galaxies themselves and
the stars in them.  How this proceeds and what are the physical
links between both SMBH and star formation remains to be
understood.

\begin{figure}[ht]
  \begin{center}
    \epsfig{file=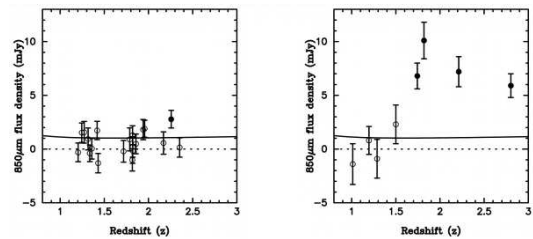, height=7cm, angle=270}
  \end{center}
\caption{Submillimeter emission for a sample of unobscured AGN
(left) and obscured AGN (right), showing that star formation is
far more important in the latter (adapted from Page et al.
2004)).\label{xbarcons_fig7}}
\end{figure}

Phenomenological links between SMBH growth and star formation have
been found over the last years by comparing X-ray fluxes to
submillimeter observations.  The first clue of co-eval SMBH growth
and star formation in AGN was reported by \cite*{Page01}, where it
was found that half of the X-ray emitting AGN were also strong
submillimeter emitters, implying high star formation rates. It is
also known that at least $\sim 40\%$ of star-forming submillimeter
emitting galaxies contain a growing SMBH emitting X-rays.
\cite*{Page04} (and see Fig.~\ref{xbarcons_fig7}) find that star
formation, as revealed by submillimeter emission, is much stronger
in obscured accreting SMBHs than in unobscured QSO-type X-ray
emitting AGN.  This suggests that SMBHs undergo a growth phase
co-eval with copious star formation, then they shine as type 1
unobscured AGN and when there is no more material to accrete SMBHs
stay dormant in galactic centers as we see most of them today.

\begin{figure}[ht]
  \begin{center}
    \epsfig{file=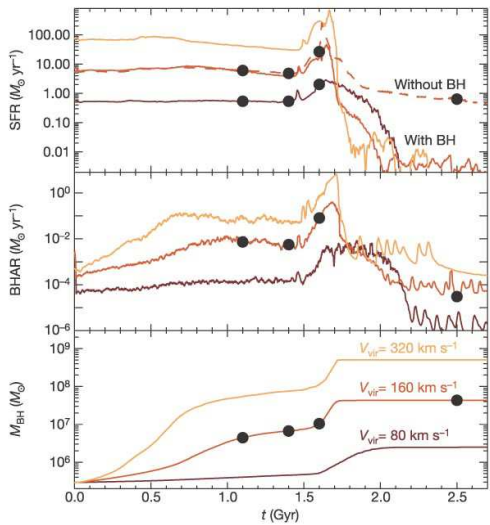, height=7cm, angle=0}
  \end{center}
\caption{Star formation rate (top), SMBH accretion rate (middle)
and SMBH mass (bottom) as a function of time in a galaxy-galaxy
collision. The top panel also illustrates what is the effect of
the SMBH accretion (adapted from Di Matteo et al 2005).
\label{xbarcons_fig8}}
\end{figure}

\cite*{DiMatteo05} have recently conducted simulations that
simultaneously follow star formation and the growth of black holes
during galaxy-galaxy collisions. In the collision there is a burst
of star formation, and large amounts of gas are funneled to the
SMBHs leading to copious accretion. The energy released ends up
expelling gas and preventing further star formation and SMBH
growth after a short phase of 100 million years (see
Fig.~\ref{xbarcons_fig8}). \cite*{DiMatteo05} also compute the
star formation rate in the absence of SMBH, leading to a much
weaker peak during the merging epoch but with a sustained rate
after the merging.  It is then clear that SMBH accretion gives
first a burst of star formation, but after the collision it
suppresses star formation.

To properly test these models and to witness how SMBH growth and
star formation in galaxies are related along cosmic history, deep
X-ray surveys need to be combined with deep surveys in the
far-infrared. Space observatories operating at these wavelengths
need to be priorities in Cosmic Vision 2015-2025.

\section{What tools are needed?} \label{sec:future}

Previous sections present a number of very exciting questions
about the energetic Universe that are being put forward by
scientists working in the field.  These questions can be
summarized as:
\begin{itemize}

\item How does matter behave under very strong
gravitational fields or at supra-nuclear densities?

\item How do baryons assemble into cosmic structures? Where are
all missing baryons gone? When, how and where were the heavy
elements present in today's Universe produced?

\item How do supermassive black holes grow, what are their parent
seeds and how are black hole growth and star formation related?

\end{itemize}

The main space tool urgently needed in the 2015-2025 decade to
address most of these questions is a large aperture (effective
area $\sim 10\, {\rm m}^2$ at 1 keV), high angular resolution ($<
5"$) X-ray observatory.  This telescope should be equipped with a
payload complement on its focal plane that allows scientists to
conduct large field-of-view deep imaging, high spectral resolution
spectroscopy at X-ray energies from 0.2 to 8 keV, a detector that
can handle large count rates from bright sources to perform timing
analysis, and a facility that extends the performance of this
telescope system beyond 30 keV. In the paper by E. Costa et al.
(these proceedings) the advantages of adding a polarimeter are
highlighted. With such an observatory in operation, the vast
majority of the scientific questions raised in this paper sould
find an answer.

ESA and JAXA have been studying a mission called XEUS (X-ray
Evolving Universe Spectroscopy
mission\footnote{http://www.rssd.esa.int/XEUS}) for a number of
years, in an attempt to fulfill the above requirements. NASA has
also been studying a mission called {\it
Constellation-X}\footnote{http://constellation.gsfc.nasa.gov} with
a special emphasis on spectroscopy, but that could also deliver
some of the science discussed in the present paper. Clearly a way
ahead should be found that secures that soon after 2015 there is a
large X-ray observatory-class space facility in operation.

The next tool obviously needed is a gravitational wave observatory
such as LISA.  This will help to detect steep variations of strong
gravity fields (those produced by BHs) out to much larger
distances than any electromagnetic radiation detector can reach.
As it has been said in subsection \ref{subsec:gravity}, the
ultimate probe of the structure of spacetime at the event horizon
itself can only be addressed by gravitational wave observatories.
An imaging $\gamma$-ray observatory will also be of enormous help
in looking at the regions next to the event horizon as well as to
detect traces of the rarest heavy elements via nuclear lines.

Next in the list, a far infrared observatory with enough
sensitivity (i.e., effective area and angular resolution) that can
trace star formation rates in obscured objects out to the
redshifts where the first galaxies, along their stars and black
holes, formed.  In combination with the large aperture X-ray
observatory, this will help us to understand the link between
supermassive black hole birth and growth and star formation.

Last, but not least, a dedicated mission to survey large parts of
the Universe with enough sensitivity (similar to DUET, DUO,
LOBSTER or ROSITA) would find the most extreme objects in the
Universe, build complete samples of rare energetic objects (such
as luminous clusters of galaxies at early epochs) and will
ultimately permit the use of galaxy clusters as cosmological
tools.

Too much, perhaps, for a single decade and meagre budget. However,
the scientific challenge and excitement is there and will not
disappear.

\begin{acknowledgements}

I'm grateful to many colleagues for help in the preparation of
this presentation and for long-standing collaborations in the
field: M. Arnaud, D. Barret, G. Bignami, J. Bleeker, F. Carrera,
A. Comastri, A.C. Fabian, G. Hasinger, H. Inoue, H. Kunieda, J.-W.
den Herder, M. M\'endez, G. Palumbo, A. Parmar, M. Turner and  C.
Turon. Financial support was provided by the Spanish Ministerio de
Educaci\'on y Ciencia, under project ESP2003-00812.

\end{acknowledgements}

\end{document}